\documentclass[a4paper,11pt]{article}
\pdfoutput=1 % if your are submitting a pdflatex (i.e. if you have
\usepackage{jcappub} % for details on the use of the package, please
\usepackage[T1]{fontenc} % if needed
\usepackage{BOONDOX-cal}
\usepackage{soul}
\usepackage{subfig}
\usepackage{gensymb}

\newcommand{\be}{\begin{equation}}
\newcommand{\ee}{\end{equation}}
\newcommand{\bea}{\begin{eqnarray}}
\newcommand{\eea}{\end{eqnarray}}

\newcommand{\Fermi}{{\it Fermi}}
\newcommand{\Jf}{${J}$}

\def\GravSphere{{\sc GravSphere}}
\def\siglos{\sigma_{\text{\rm l.o.s.}}}

\def\vlosfour{\langle v_{\rm l.o.s.}^4 \rangle}
\def\vsone{v_{s1}}
\def\vstwo{v_{s2}}

\title{\boldmath Dark matter constraints from dwarf galaxies with data-driven J-factors}

\author[a,b]{Alexandre Alvarez,}
\author[b]{Francesca Calore,}
\author[c]{Anna Genina,}
\author[d]{Justin Read,}
\author[b]{Pasquale Dario Serpico,}
\author[e]{and Bryan Zaldivar}

\affiliation[a]{Institut f\"{u}r Theoretische Physik und Astrophysik, Uni W\"{u}rzburg, Germany}
\affiliation[b]{Univ. Grenoble Alpes, USMB, CNRS, LAPTh, F-74940 Annecy, France}
\affiliation[c]{Institute for Computational Cosmology, Department of Physics, Durham University, South Road, Durham DH1 3LE, UK}
\affiliation[d]{Department of Physics, University of Surrey, Guildford, GU2 7XH, UK}
\affiliation[e]{Dpto. and Instituto de Fisica Teorica, IFT-UAM/CSIC, Cantoblanco, 28049, Madrid, Spain}
\emailAdd{alexandre.alvarez@physik.uni-wuerzburg.de}
\emailAdd{calore@lapth.cnrs.fr}
\emailAdd{anna.genina@durham.ac.uk}
\emailAdd{j.read@surrey.ac.uk}
\emailAdd{serpico@lapth.cnrs.fr}
\emailAdd{bryan.zaldivarm@uam.es}

\abstract{We present an updated analysis of the gamma-ray flux from the directions of classical dwarf spheroidal galaxies, deriving new constraints on WIMP dark matter (DM) annihilation using a decade of \Fermi-LAT data. Among the major novelties, we infer the dwarfs' \Jf-factors by including new observations without imposing any {\it a priori} parametric profile for the DM distribution. While statistically compatible with results obtained from more conventional parameterisations, this procedure reduces the theoretical bias imposed on the data. Furthermore, we retain the full data-driven shape of the \Jf-factors' empirical probability distributions when setting limits on DM, without imposing log-normality as is typically done. In conjunction with the data-driven \Jf-factors, we improve on a new method for estimating the probability distribution function of the astrophysical background at the dwarf position~\cite{Calore_2018}, fully profiling over background uncertainties. We show that, for most ``classical'' dwarfs, the background systematic uncertainty dominates over the uncertainty on their \Jf-factors. Raw distributions of \Jf- and $\mathcal{D}$-factors (the latter being the analogous of \Jf-factors for decaying DM) are available upon request.}

\begin{document}

\begin{flushright}
LAPTH-002/20\\
IFT-UAM/CSIC-20-15
\end{flushright}

\maketitle
\flushbottom

\section{Introduction}
\label{introd}
Over the last few decades, our understanding of the non-thermal universe has greatly advanced thanks to observational progresses and the introduction of more refined analysis techniques. This progress has deepened our understanding of astrophysical objects and environments, but it has also increased our ability to search for hypothetical physical phenomena, such as dark matter (DM) annihilation. Annihilation byproducts are expected to be detectable in leading particle physics theories for DM (so-called WIMP scenarios), and can shed light on the nature of this species which till now has only been detected gravitationally, see, e.g.,~\cite{Bertone:2010zza}.

A paradigmatic case is the gamma-ray one: currently, we understand that most of the detected photons are either diffuse photons coming from elementary processes involving cosmic rays in the interstellar medium, or emitted by powerful astrophysical sources such as active galactic nuclei, pulsars and their nebulae, or supernova remnants~\cite{2011hea..book.....L}. As a result, DM searches are definitely not ``background free'', and their sensitivity reach is limited not only by statistics, but increasingly by the systematic error associated with our understanding and modelling of competing, and typically dominant, astrophysical emissions~\cite{Charles:2016pgz}. In this context, the most promising targets in the energy range probed by the Large Area Telescope, aboard the \Fermi~satellite (\Fermi-LAT), is the gamma-ray flux expected from DM-rich satellites of the Milky Way, known as dwarf spheroidal galaxies (dSphs), as has been recognised for a long time~\cite{Lake:1990du,Evans:2003sc}. Since their intrinsic gamma-ray background is negligible given current instrumental sensitivity~\cite{Winter:2016wmy}, searches mostly suffer from diffuse backgrounds (and, possibly, other gamma-ray sources) that lie along the line of sight (l.o.s.).

Anticipating that this will eventually become a crucial limitation, in a recent article~\cite{Calore_2018} some of us proposed a data-driven method to estimate this astrophysical background. This consists of a probability density estimation by a kernel method, where the probability distribution function (PDF)
of the photon counts in the ``signal'' direction is inferred from the counts in the pixels in a surrounding area, whose size and weights are optimised purely based on the data. 
The performance is typically comparable with more traditional techniques, but one can more neatly account for the background uncertainty in a statistically meaningful way.

In this article, whose structure we briefly summarise below, we move some steps further along this direction. After recalling some basic notation and the data sample used in Sec.~\ref{generalities}, in Sec.~\ref{sampleJ} we describe  the new determination of the DM signal strengths, depending on the so-called astrophysical ``$J$-factors". For these, we follow the same rationale as for the background estimate, namely: We adopt a data-driven, non-parametric procedure to estimate them, without imposing a priori a DM density profile form. Furthermore, we implement a few technical improvements in the procedure to estimate the astrophysical background, and perform additional tests of its robustness, as described in Sec.~\ref{bkg}. We present the limits on the DM annihilation parameter space in Sec.~\ref{results}. Our conclusions are reported in Sec.~\ref{conclusion}. In Appendix~\ref{sec:Dfactors} we present results for the ``$\mathcal{D}$-factors'', the analogous of \Jf-factors relevant for decaying DM signals.

\section{Generalities, basic notations and data treatment}\label{generalities}

Here we recap the basic notation and formalism, addressing the reader to the previous publication~\cite{Calore_2018} for further details. 
The essential astrophysical input for DM annihilation searches from dSphs are the so-called \Jf-factors, defined as~\cite{Bringmann:2012ez}: 
\begin{equation}
    J(\Delta \Omega) = \int_{\Delta \Omega} \int_{\rm l.o.s} \rho^2 (l,\Omega') {\rm d}l {\rm d}\Omega' .
    \label{eq:J_factor}
\end{equation}
In the equation above, $l$ is the l.o.s. coordinate, $\rho$ is the radius-dependent DM density of the dSph under consideration and $\Delta \Omega$ is the solid angle over which we integrate, fixed henceforth to a circle of 0.5$^\circ$ radius. The \Jf-factors enter the differential gamma-ray flux ${\rm d}\phi/{\rm d}E$ from each dSph as a multiplicative factor:
\begin{equation}
    \frac{{\rm d} \phi}{{\rm d}E} = \frac{1}{4\pi}\frac{\langle \sigma v\rangle J(\Delta \Omega)}{2m^2_{DM}}\frac{{\rm d}N}{{\rm d}E}\,,
    \label{eq:flux_diff}
\end{equation} 
where ${\rm d}N/{\rm d}E$ is the differential number of photons per annihilation (for the specific channel considered) taken from~\cite{Cirelli:2010xx}, $m_{DM}$ is the mass of the DM particle (assumed self-conjugated, hence the factor 1/2), and $\langle \sigma v \rangle$ is the velocity-averaged annihilation cross section. We assume annihilation to $b\bar b$ throughout. 
The estimated photon count number $ \lambda^{DM}_{d,e}$ for a dSph (indexed by $d$) in a given energy bin (denoted by $e$) requires the convolution of the flux in equation \eqref{eq:flux_diff}, above, with the exposure map at the position of the dSph, $\mathcal{E}_d (E)$: 
\begin{equation}
    \lambda^{\rm DM}_{d,e} = \int_e \frac{{\rm d}\phi}{{\rm d} E} \mathcal{E}_d (E){\rm d}E = J_d \langle \sigma v \rangle f_{d,e} (m_{DM}) \,.
    \label{eq:count_DM}
\end{equation}
The $f_{d,e} (m_{DM})$ function takes into account the spectral information of the signal and the convolution with the exposure map. 

Of course, the total number of photons $\lambda_{d,e}$ one measures also includes -- and is actually dominated by -- background photons, such that:
\begin{equation}
    \lambda_{d,e} = \lambda^{\rm DM}_{d,e} + b_{d,e} \simeq b_{d,e} ,
\end{equation}
with $b_{d,e}$ being the number of photons from the background for the dSph $d$ in the energy bin $e$. 
The whole theoretical problem is thus reduced to an estimation of the (probability distributions of) \Jf-factors and  the background, which are addressed in the two following sections, respectively.

Another relatively straightforward update concerns the larger gamma-ray statistics, which now spans almost a decade of \Fermi-LAT observations.
We select data taken from the LAT from week 9 up to week 522 (from August 4, 2008 to June 7, 2018).
We select events from the SOURCE \Fermi-LAT P8R2 class\footnote{Note that we choose the SOURCE event class, rather than the CLEAN class, to avoid being  too much penalised by the low statistics at high-energy.}, both front- and back-converted, spanning the energy interval 500 MeV $-$ 500 GeV.  
To prepare the data sample, we apply standard cuts (``(DATA\_QUAL$>$0) \&\& (LAT\_CONFIG==1)'' and zenith angles $> 90^\circ$). 
The all-sky data are available in 24 logarithmically spaced energy bins and in angular pixels (Cartesian projection) of 0.1$^\circ \times$0.1$^\circ$ solid angle size. For the actual analyses, we re-bin them in six logarithmically spaced energy bins and in angular bins of $0.5^\circ$ radius, as done in~\cite{Calore_2018}.

\section{New \Jf-factors determination}\label{sampleJ}

We determined the \Jf-factors using the \GravSphere\ method described in \cite{Read17}, \cite{Read18} and \cite{Genina19}. This is an non-parametric Jeans mass modelling method \cite{1922MNRAS..82..122J} that assumes a dynamic steady state and spherical symmetry, solving the following system of equations:

\begin{equation}
\siglos^2(R)= \frac{2}{\Sigma(R)}\int_R^\infty \left(1\!-\!\beta\frac{R^2}{r^2}\right)
    \nu\sigma_r^2\,\frac{r\,dr}{\sqrt{r^2\!-\!R^2}} \ ,
    \label{eqn:LOS}
\end{equation}
where $\Sigma(R)$ is the tracer surface mass density at a projected radius, $R$, and $\nu(r)$ is the 3D tracer density at spherical radius, $r$,
\begin{equation}
\sigma_r^2(r) = \frac{1}{\nu(r) g(r)} \int_r^\infty \frac{GM(\tilde{r})\nu(\tilde{r})}{\tilde{r}^2} g(\tilde{r}) \, \tilde{r}
\label{eqn:main}
\end{equation}
is the radial velocity dispersion,
\begin{equation}
g(r) = \exp\left(2\int_0^r \frac{\beta(\tilde r)}{\tilde r}\, \tilde r\right),
\label{eqn:ffunc}
\end{equation}
and
\begin{equation}  \label{beta}
\beta = 1 - \frac{\sigma_t^2}{\sigma_r^2}
\end{equation}
is the velocity anisotropy parameter, where $\sigma_t$ is the tangential dispersion. Finally, $G$ is Newton's gravitational constant, and $M(r)$ is the cumulative mass of the stellar system (due to all stars, gas, DM etc.) that we would like to measure.

By default, \GravSphere\ also fits the two higher-order ``Virial Shape Parameters'' (VSPs; \cite{1990AJ.....99.1548M,2014MNRAS.441.1584R}):

\begin{eqnarray} 
\vsone & = & \frac{2}{5}\, \int_0^{\infty} G M\, (5-2\beta) \,\nu \sigma_r^2 \,r \,dr \\
\label{eqn:vs1}
& = & \int_0^{\infty} \Sigma \vlosfour\, R\, dR
\label{eqn:vs1data}
\end{eqnarray}
and:
\begin{eqnarray} 
\vstwo & = & \frac{4}{35} \,\int_0^{\infty} G M\, (7-6\beta)\, \nu \sigma_r^2 \,r^3 \,dr \\
\label{eqn:vs2}
& = & \int_0^{\infty} \Sigma \vlosfour\, R^3\, dR \ .
\label{eqn:vs2data}
\end{eqnarray}
where $\vlosfour$ is the fourth moment of the l.o.s.~velocities. This breaks the otherwise trivial degeneracy between $\beta(r)$ and the cumulative mass profile $M(r)$ \cite{1990AJ.....99.1548M,2014MNRAS.441.1584R,Read17}.

The \Jf-factor then follows from a numerical double integral of the DM density, that is the radial derivative of the DM component of $M(r)$ (see equation \ref{eq:J_factor}).
We solve the above equations to obtain \Jf-factors for  eight classical Milky Way dSphs and Segue I by fitting their l.o.s.~velocity dispersion, $\siglos$, tracer surface density, $\Sigma(R)$, and VSPs as in \cite{Read19}. For the ultra-faint dSph Segue I, we use the data described in \cite{Simon11}, with membership probability 0.9 or 0.95, as in \cite{Bonnivard16}.
For the fits, we use both the \texttt{FreeForm} mass model described in \cite{Read17}, and the \texttt{coreNFWtides} model described in \cite{Read18}, with priors as in \cite{Read19b}.

\begin{figure}[!h]
  \centering{
  \subfloat{\includegraphics[width=5.2cm]{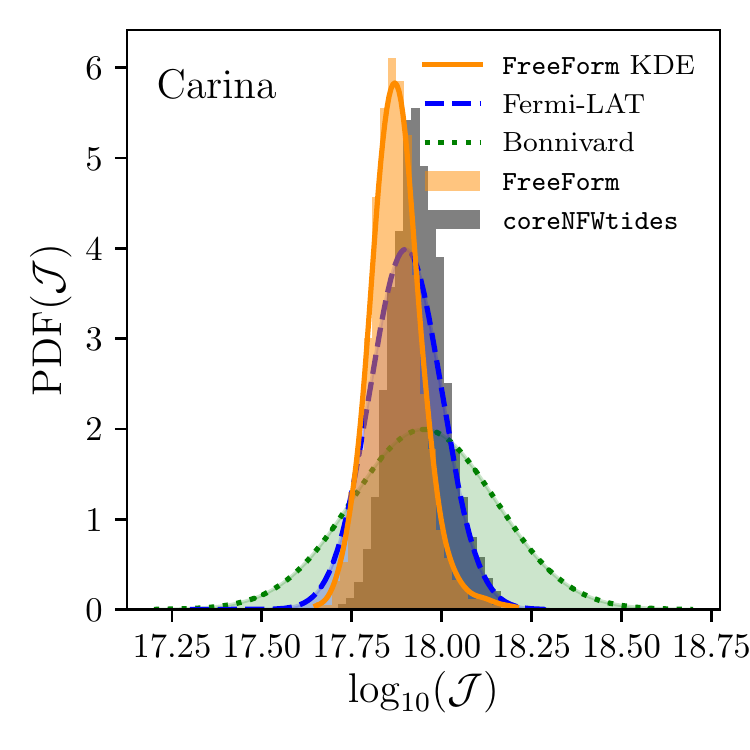}}
  \subfloat{\includegraphics[width=5.2cm]{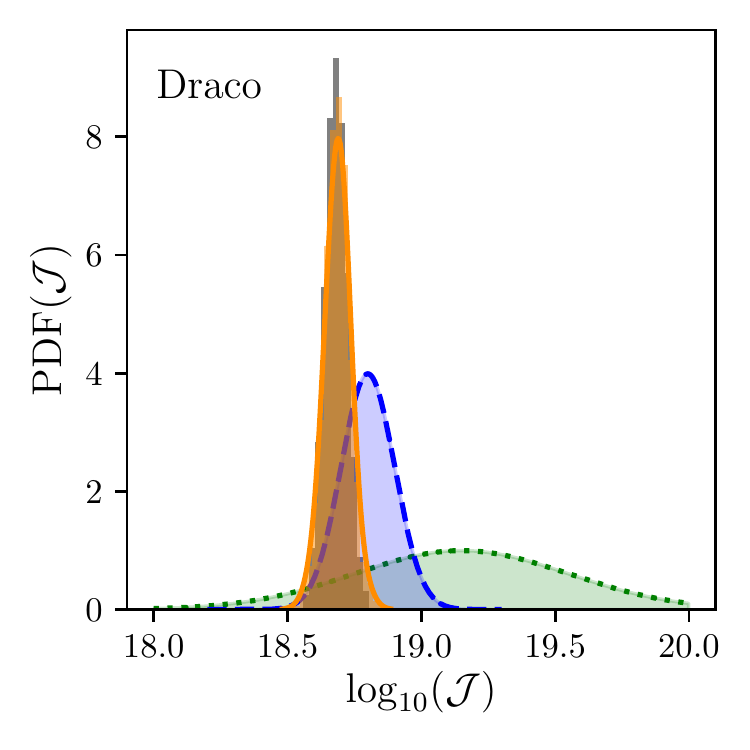}}
  \subfloat{\includegraphics[width=5.2cm]{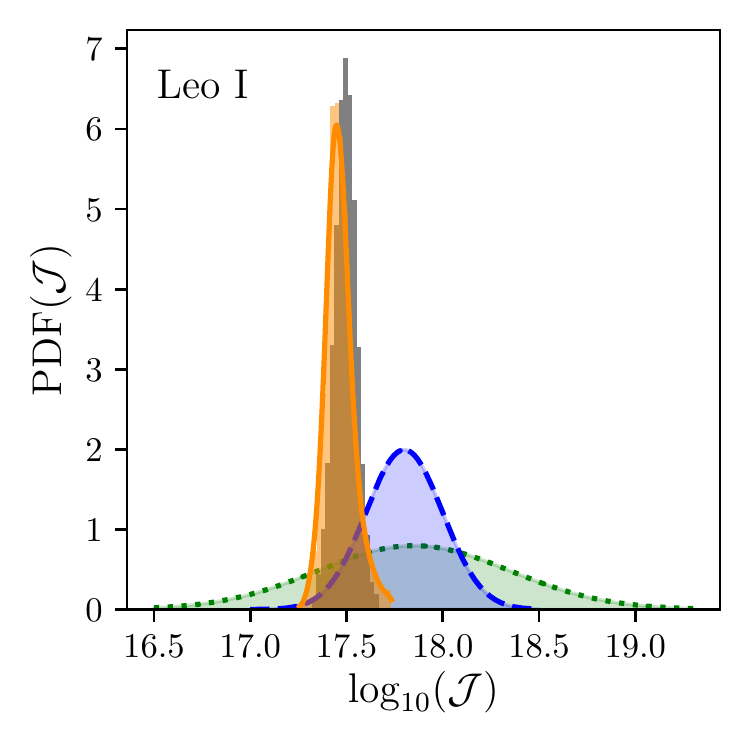}}\\
  \subfloat{\includegraphics[width=5.2cm]{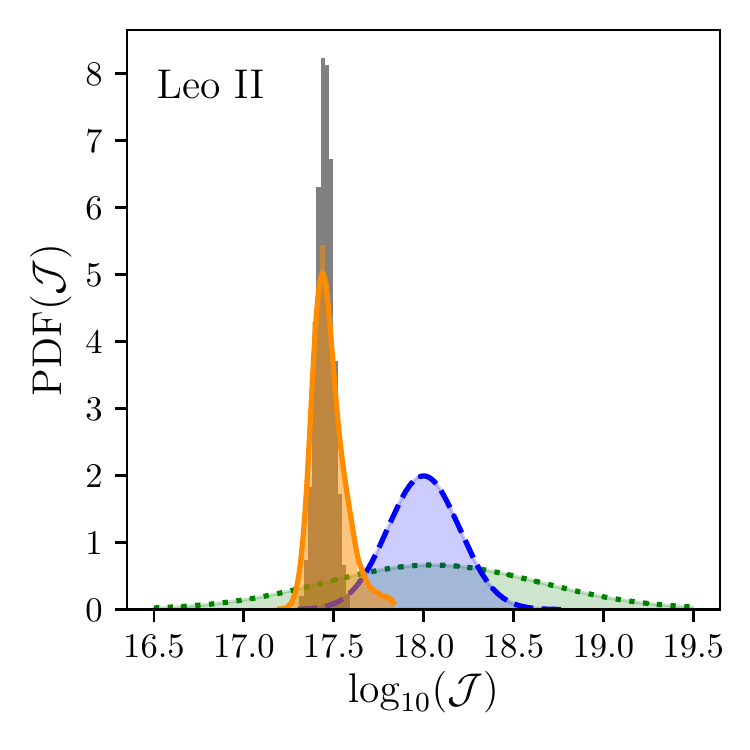}}
  \subfloat{\includegraphics[width=5.2cm]{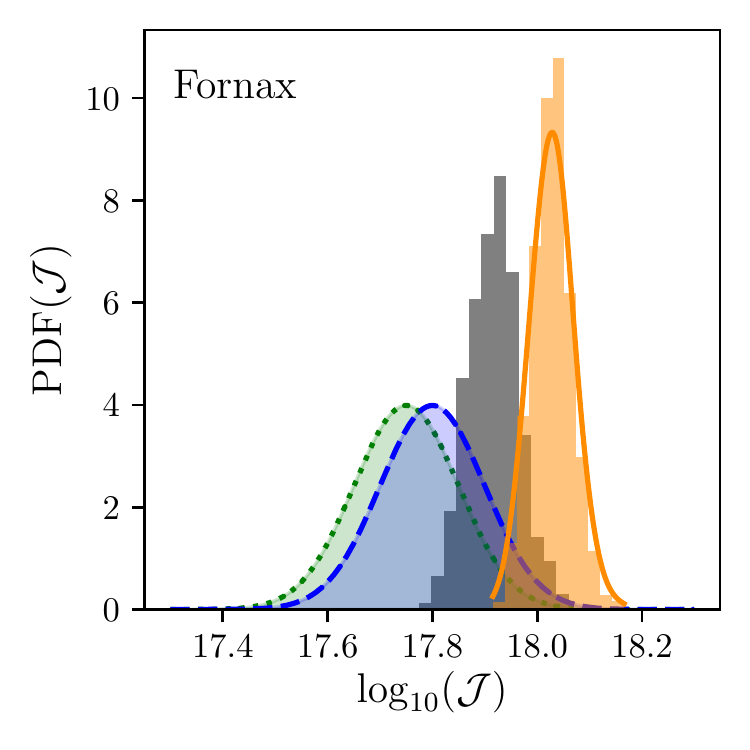}}
  \subfloat{\includegraphics[width=5.2cm]{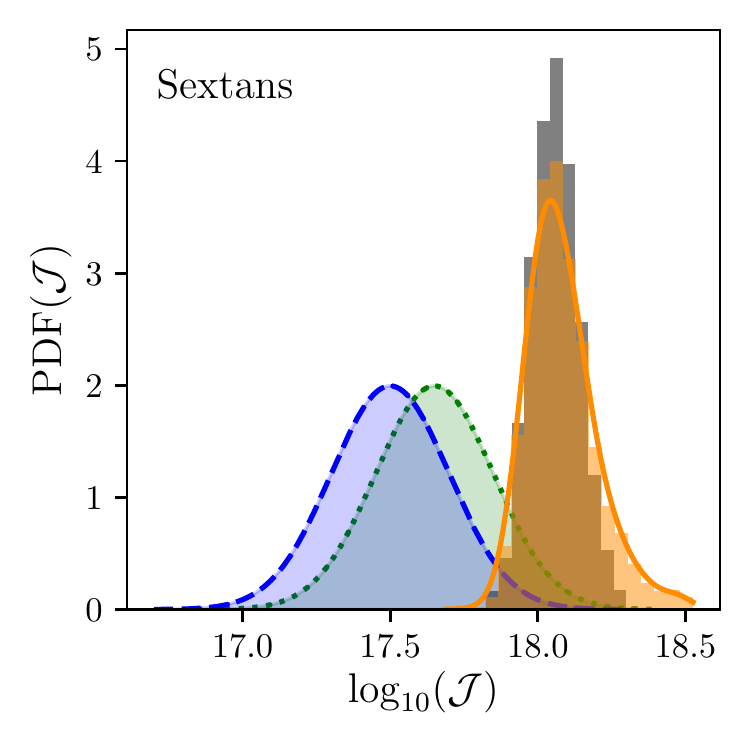}}\\
  \subfloat{\includegraphics[width=5.2cm]{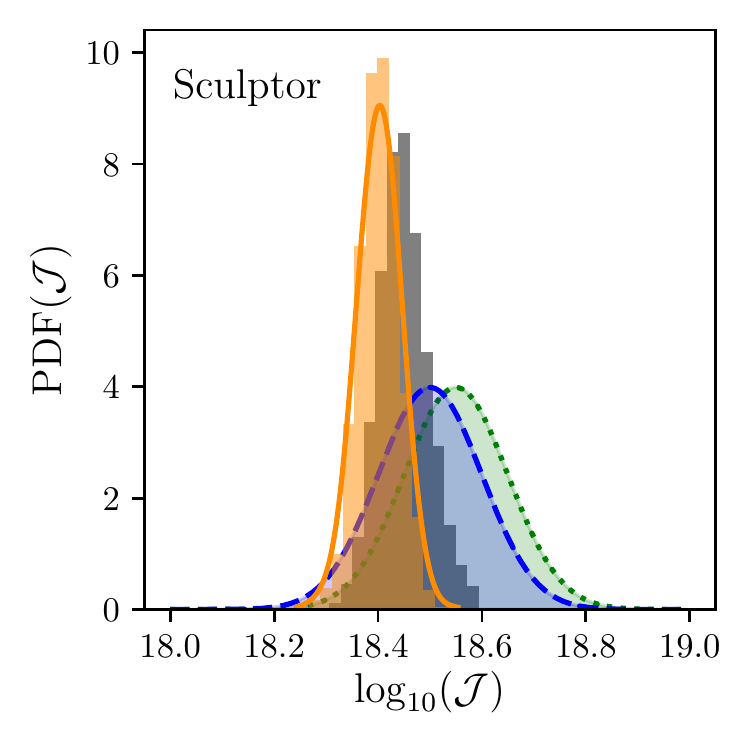}}
  \subfloat{\includegraphics[width=5.2cm]{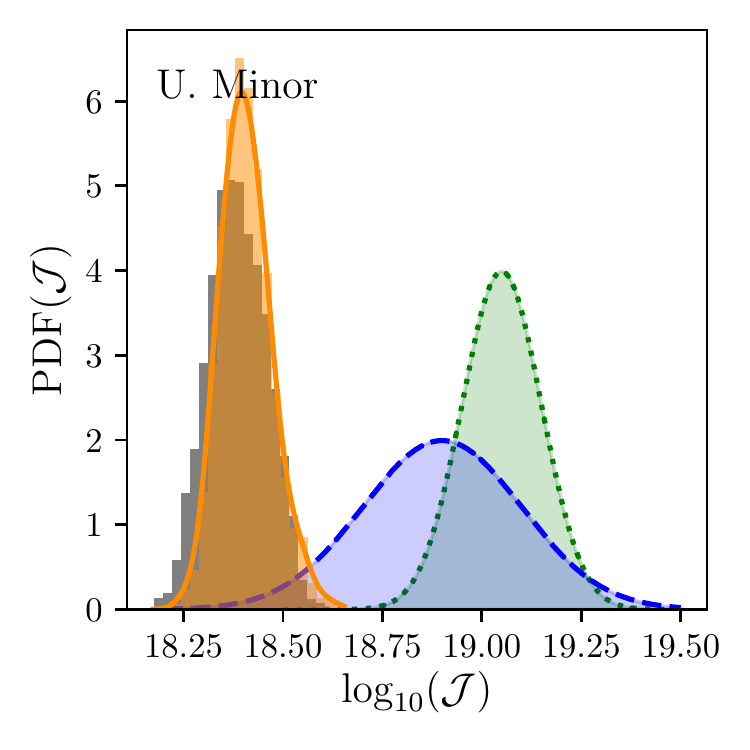}}}
  \caption{New \Jf-factor distributions derived for the 8 ``classical'' dSphs in the \texttt{FreeFrom} case (orange histogram) compared with the  \texttt{coreNFWtides} case (grey histograms), and the log-normal distributions commonly used (e.g. by the \Fermi-LAT collaboration~\cite{Fermi-LAT:2016uux} and in our previous article~\cite{Calore_2018}, dashed-blue), as well as the ones from~\cite{Bonnivard:2015xpq} (dotted-green).}
\label{fig:J-factors_new}
\end{figure}

\begin{figure}[!ht]
  \centering{
  \subfloat{\includegraphics[width=5.5cm]{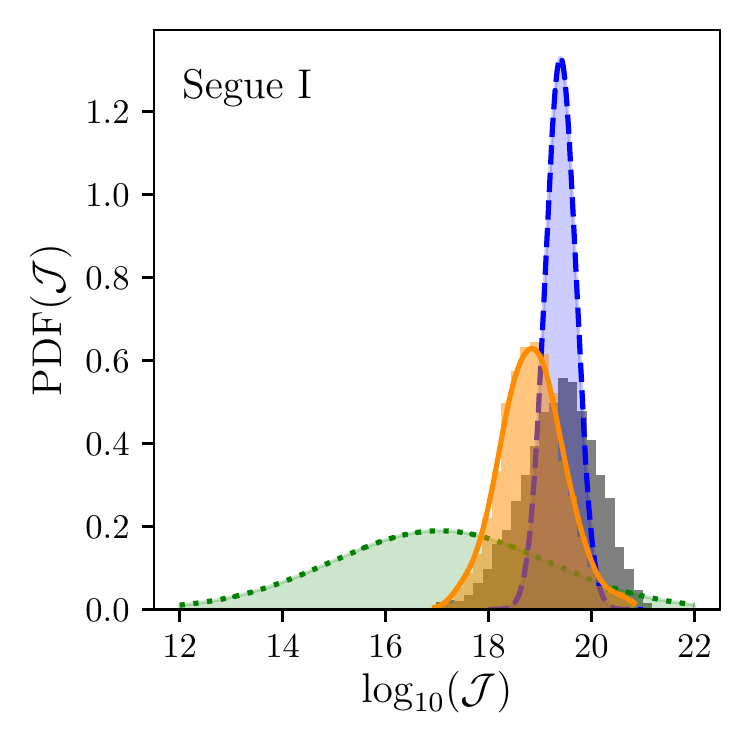}}
  \subfloat{\includegraphics[width=5.5cm]{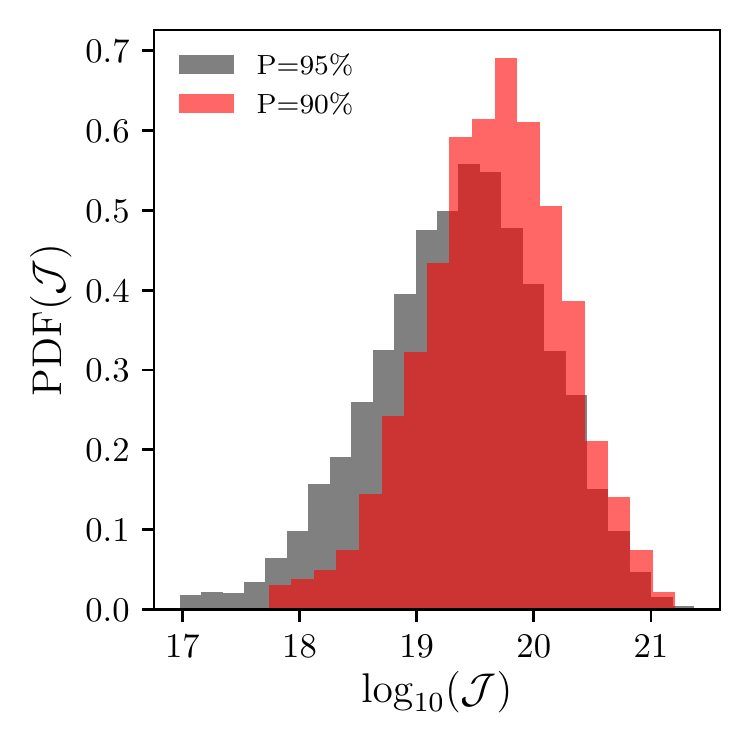}}
  \caption{\emph{Left}: Same as Fig.~\ref{fig:J-factors_new} for Segue I. \emph{Right}:  \texttt{coreNFWtides} model \Jf-factor PDF for Segue I using different membership probability cuts (90\% and 95\%, as marked).}\label{fig:J-factors_segue}}
\end{figure}

 Fig.~\ref{fig:J-factors_new} shows the \Jf-factor distributions corresponding to the newly derived DM density profiles for eight classical dSphs in the \texttt{FreeFrom} case (orange histogram) compared with the \texttt{coreNFWtides} case (grey histograms), and the log-normal distributions commonly used (e.g. by the \Fermi-LAT collaboration~\cite{Fermi-LAT:2016uux} and in our previous article~\cite{Calore_2018}, dashed-blue), as well as the ones from~\cite{Bonnivard:2015xpq} (dotted-green).
Overall, we find a good agreement with existing determinations of the classical dSphs, with the largest deviation arising for Sextans and U. Minor. This can be traced to the rather different half-light radii that Read et al.~\cite{Read19b} find for these dwarfs as compared to literature values (see their Tab.~1). In most cases, the new PDFs are narrower, which implies a reduced contribution of the \Jf-factor uncertainties to the total error budget. Notice that the \texttt{coreNFWtides} and \texttt{FreeForm} results overlap well, with the largest deviation for Fornax being still only at the one sigma level. This is most likely because Fornax is the most cored of all the dSphs in our sample \citep{Read19b}, while the \texttt{FreeForm} model and priors favour cusps over cores in the absence of data \citep{Read18}. Notice further that the \texttt{coreNFWtides} determinations are characterised by slightly narrower distributions, as expected due to its more constrained underlying functional form.
 
 It is reassuring that the choice of mass model does not induce a bias that is larger than our quoted uncertainties. In the following, we will use smoothed versions of the \texttt{FreeFrom} determination, with a smoothing parameter that is typically one order of magnitude smaller than the width of the distribution. We will make sure that the results are not sensitive to this choice.
A log-normal distribution, whose parameters for each dSph are quoted in the third column of Tab.~\ref{tab:pred}, provides a good approximation to all of the dSphs in our sample, with differences on the bounds at the tens of percent level, at most.

Finally, in Fig.~\ref{fig:J-factors_segue} we show our results for the \Jf-factor determination for Segue I. Although the average \Jf-factor determination is higher than the classical dSphs considered here, its uncertainty is one order of magnitude larger. We can thus expect that its constraining power will not be the best, once profiling over its uncertainties. Note, however, that the systematic uncertainties due to the choice of membership probability cut (right panel in Fig.~\ref{fig:J-factors_segue}) or of the choice of mass model (\texttt{coreNFWtides} vs. \texttt{FreeForm}; grey vs.~orange histograms in the left panel of Fig.~\ref{fig:J-factors_segue}) are below one sigma. This leaves significant room for improvement in reducing statistical uncertainties through appropriate observational campaigns targeting ultra-faint galaxies like Segue I.

With the same dSph mass profile we can also compute $\mathcal{D}$-factors, i.e. the integral along the l.o.s. of the DM density, and relevant for predictions of decaying DM signals. We present the newly determined $\mathcal{D}$-factors in Appendix~\ref{sec:Dfactors} for the  \texttt{FreeForm} model, and we also compare them with a previous determination from~\cite{Geringer-Sameth:2014yza}.

\section{Background estimation: Formalism and novelties}
\label{bkg}
Overall, we follow the same non-parametric method detailed in \cite{Calore_2018} for the determination of the background at the positions of dSph, as well as its PDF. Our results (for the energy-integrated background, although the energy-dependent one is used in actual calculations) are summarised in Tab.~\ref{tab:pred}.
Below, we recall briefly the main ingredients entering our estimate. For the definition of the likelihood function $F$, we  adopt a kernel estimate of the PDF of the output $b$, the background counts:\\
\begin{eqnarray}
    F(\Vec{x},b) &&= \frac{1}{N} \sum_{i=1}^{N} K_{\sigma}(\Vec{x} - \Vec{x_i}) \times g_{\varsigma}(b, b_i)=\\
    &&= \frac{1}{N} \sum_{i=1}^{N} \frac{1}{2 \pi \sigma^2 } \text{exp} \left(- \frac{(\Vec{x} - \Vec{x_i})^T(\Vec{x}-\Vec{x_i})}{2\sigma^2}\right) \times \frac{1}{\sqrt{2 \pi} \varsigma b} \text{exp}\left(-\frac{(\text{ln}(b) - \text{ln}(b_i))^2}{2\varsigma ^2}\right)\,, \nonumber
\label{F_likelihood}
\end{eqnarray}
where the position kernel $K_{\sigma}$ and the count kernel $g_{\varsigma}$, in the second step, have been chosen to be a Gaussian and a log-normal, respectively, and $\Vec{x}$ is a short-hand for the coordinates determining the direction in the sky. 
We then construct the total Likelihood across all the control regions as
\begin{equation}
    F_{\rm tot}(\sigma, \varsigma) = \prod_{i=1}^N \left( \frac{1}{N-1}\sum_{j \neq i}^{N-1}  K_{\sigma}(\Vec{x_i} - \Vec{x_j})g_{\varsigma}(b_i, b_j)\right) 
    \label{kernel}
\end{equation}
and use it for our Maximum Likelihood Estimation (MLE) of the parameters $\sigma$, $\varsigma$, by following a grid-search optimisation method. As a result, we obtain the optimal parameters
$\sigma^*$ and $\varsigma^*$ defined as
\begin{equation}
    (\sigma^*,\varsigma^*) = \text{argmax ln }( F_{\rm tot}(\sigma, \varsigma)) .
\label{eq:sig_varsig}
\end{equation}

We introduce three novelties based on the insights gained from our previous analysis: i) We apply a simplification in the generation of the background pixels, described in Sec.~\ref{isotropic}.
ii) We check the impact of optimising further the background estimate, by allowing for multiple (energy-dependent) hyper-parameters (Sec.~\ref{E-dependent}). iii) We study the impact of relaxing somewhat the hypothesis of strongly correlated background in different energy bins, as described in Sec.~\ref{bin_correlation}. 
\begin{table}
\begin{center}
\begin{tabular}{| c | c | c | c | c | c |}
\hline
dSph & name & $\log_{10} J \pm \sigma^J$ & $c \pm \sqrt{c} $ & 
$\tilde{b}$ & $\widehat{\ln\sf b} \pm \Delta(\ln{\sf b})$ \\
\hline\hline
1& Bo\"otes I & -- & 102.0  $\pm$  10.0  &  112.52  &  4.72  $\pm$  0.16  \\
\hline
2 & Canes Venatici I & -- & 27.0  $\pm$  5.0  &  38.88  &  3.66  $\pm$  0.23\\
\hline
3 & Canes Venatici II & -- & 27.0  $\pm$  5.0  &  25.83  &  3.25  $\pm$  0.27 \\
\hline
4 & Carina & $17.87 \pm 0.07$ & 336.0  $\pm$  18.0  &  334.48  &  5.81  $\pm$  0.17 \\
\hline
5 & Coma Berenices & -- & 39.0  $\pm$  6.0  &  34.78  &  3.55  $\pm$  0.36 \\
\hline
6 & Draco$^{*}$ & $18.69 \pm 0.05$ & 267.0  $\pm$  16.0  &  287.37  &  5.66  $\pm$  0.21 \\
\hline
7 & Fornax & $18.03 \pm 0.04$ & 100.0  $\pm$  10.0  &  89.67  &  4.5  $\pm$  0.2  \\
\hline
8 & Hercules & -- & 450.0  $\pm$  21.0  &  409.83  &  6.02  $\pm$  0.16 \\
\hline
9 & Horologium I & --  & 166.0  $\pm$  13.0  &  138.66  &  4.93  $\pm$  0.18 \\
\hline
10 & Hydra II & -- & 351.0  $\pm$  19.0  &  377.4  &  5.93  $\pm$  0.18 \\
\hline
11 & Leo I & $17.47 \pm 0.07$ & 232.0  $\pm$  15.0  &  217.0  &  5.38  $\pm$  0.16 \\
\hline
12 & Leo II$^{*}$ &$17.48 \pm 0.08$ & 82.0  $\pm$  9.0  &  96.77  &  4.57  $\pm$  0.19\\
\hline
13 & Leo IV & -- & 175.0  $\pm$  13.0  &  175.1  &  5.17  $\pm$  0.17 \\
\hline
14  & Leo V & -- & 187.0  $\pm$  14.0  &  165.51  &  5.11  $\pm$  0.18 \\
\hline
15 & Pisces II & -- & 251.0  $\pm$  16.0  &  248.08  &  5.51  $\pm$  0.19\\
\hline
16 & Reticulum II & -- & 167.0  $\pm$  13.0  &  146.76  &  4.99  $\pm$  0.17 \\
\hline
17  & Sculptor$^{*}$ & $18.40 \pm 0.04$ & 27.0  $\pm$  5.0  &  32.33  &  3.48  $\pm$  0.3 \\
\hline
18 & Segue I & $18.86 \pm 0.63$ & 211.0  $\pm$  15.0  &  194.08  &  5.27  $\pm$  0.17 \\
\hline
19 & Sextans & $18.09 \pm 0.11$ & 271.0  $\pm$  16.0  &  235.58  &  5.46  $\pm$  0.18\\
\hline
20& Tucana II & -- & 167.0  $\pm$  13.0  &  139.16  &  4.94  $\pm$  0.16\\
\hline
21  & Ursa Major I &-- & 145.0  $\pm$  12.0  &  129.26  &  4.86  $\pm$  0.19 \\
\hline
22 & Ursa Major II &-- & 369.0  $\pm$  19.0  &  358.5  &  5.88  $\pm$  0.15 \\
\hline
23 & Ursa Minor$^{*}$ &$18.41 \pm 0.06$ & 214.0  $\pm$  15.0  &  203.42  &  5.32  $\pm$  0.16\\
\hline
24 & Willman 1 & -- & 132.0  $\pm$  11.0  &  118.1  &  4.77  $\pm$  0.18\\
\hline
25 & Grus I & -- & 125.0  $\pm$  11.0  &  112.0  &  4.72  $\pm$  0.18 \\
\hline\hline
\end{tabular}
\caption{The counts (with their Poisson errors) and estimated backgrounds (with their uncertainties, in natural log units) for the directions of the 25 dSphs reported in our previous analysis~\cite{Calore_2018}. For the objects considered in this
 analysis, we further report the inferred central values and root mean square of the  
\Jf-factor (both in $\log_{10}$ scale and integrated up to 0.5$^\circ$) according to our new study. 
Asterisks mark the most significant targets we use for the combined limits.}
\label{tab:pred}
\end{center}
\end{table}
\subsection{Isotropically generated control regions}\label{isotropic}

In \cite{Calore_2018}, the control regions (i.e. non-overlapping $0.5^\circ$ radius pixels in the sky not occupied by a dSph, the Galactic plane between -20$^\circ$ and 20$^\circ$ in latitude, nor an astrophysical source from the known catalogue) used to characterise the background at the dSph positions were generated following a smoothed version of the empirical distribution of dSph, which manifests appreciable variations over tens of degrees in the sky. A posteriori, however, it turned out that the optimal background determination is controlled by a relatively small region around the dSph, with the optimal ``weight'' parameter being only slightly larger than a degree, that is the average distance to the nearest neighbours. Over this scale, the large-scale distribution of the dSphs cannot be resolved.  It is a sensible thing to test, then, if a  simplified generation procedure adopting a isotropic hypothesis is performing similarly well.

We find only modest differences in the optimization between the isotropic case or the ``empirical'' distribution case, 
\begin{equation}
    \begin{cases}
    (\sigma^*,\varsigma^*)_{\rm emp} = (1.2^\circ, 0.16)\\
    (\sigma^*,\varsigma^*)_{\rm iso} = (1.4^\circ, 0.15)
    \end{cases}
\end{equation}
with the isotropic case being slightly less noisy. For most of the dSphs, the background estimates are within the typical width of the background PDF (i.e. ``1-2$\sigma$s''). 
An isotropic distribution of void regions can instead lead to a more important improvement whenever the empirical distribution produces sparser counts in nearby pixels. Overall, given its simpler implementation and the slight improvement, we adopt in the following the isotropic case, with $N=13178$ points obeying our control region criteria. Another sanity check passed by our new procedure is that the central value of the background currently estimated is typically within the one sigma interval of the background estimated in our previous analysis~\cite{Calore_2018}, once rescaled for the different exposure time.

\subsection{Energy dependent background estimation}\label{E-dependent}
In \cite{Calore_2018}, the optimisation parameters $\sigma^*$ and $\varsigma^*$ were derived by using the data in the first energy bin only. While we argued that it is a sensible physical approximation to assume that the counts in different energy bins at the same location are heavily correlated, one may still worry that this simplified procedure may not provide an optimal estimate of the background for higher energies. For instance, higher-energy bins are less rich in counts, and it may be preferable  to average over a broader spatial region in order to reduce the ``noise'', so to speak. To assess the goodness of this hypothesis, we adopted multiple hyper-parameters to determine the background model in each energy bin.
In practice, this amounts to applying Eq.~(\ref{eq:sig_varsig}) to each energy bin. The results of this optimisation are reported in  table \ref{tab:sigma}. Although we see a slight increase of the best-fit values with energy, 
the differences among energy bins are not dramatic, with a spread of about 3\% and 10\% for $\sigma^*$ and $\varsigma^*$, respectively. More importantly, the  induced  change in the background estimation at the dSphs positions amounts to less than 1.5\%, i.e typically at least one order of magnitude below the estimated uncertainty of the background. As a consequence, we have validated the correctness of the choice to use a single set of optimisation parameters done in~\cite{Calore_2018}, which we also adopt in the following unless stated otherwise, since it is  inconsequential for the analysis.  

\begin{table}[!ht]
  \centering
  \begin{tabular}{|c|c|}
        \hline
         $E_{\rm bin}$ (GeV)& $(\sigma^*, \varsigma^*)$ \\
        \hline
          0.67 & (1.40$^\circ$,0.15)\\
        \hline
          0.89 &(1.42$^\circ$,0.16)\\
         \hline
           1.19&(1.46$^\circ$,0.18)\\
         \hline
          1.58 &(1.48$^\circ$,0.20)\\
         \hline
           2.81&(1.47$^\circ$,0.19)\\
         \hline
         500&(1.50$^\circ$,0.21)\\
         \hline
         
\end{tabular}
  \caption{Values of $\sigma^*$ and $\varsigma^*$ for every energy bin.}
\label{tab:sigma}
\end{table}

\subsection{Energy bin correlation hypothesis}\label{bin_correlation}
\begin{figure}[h]
  \centering{
  \subfloat{\includegraphics[width = 0.5\textwidth]{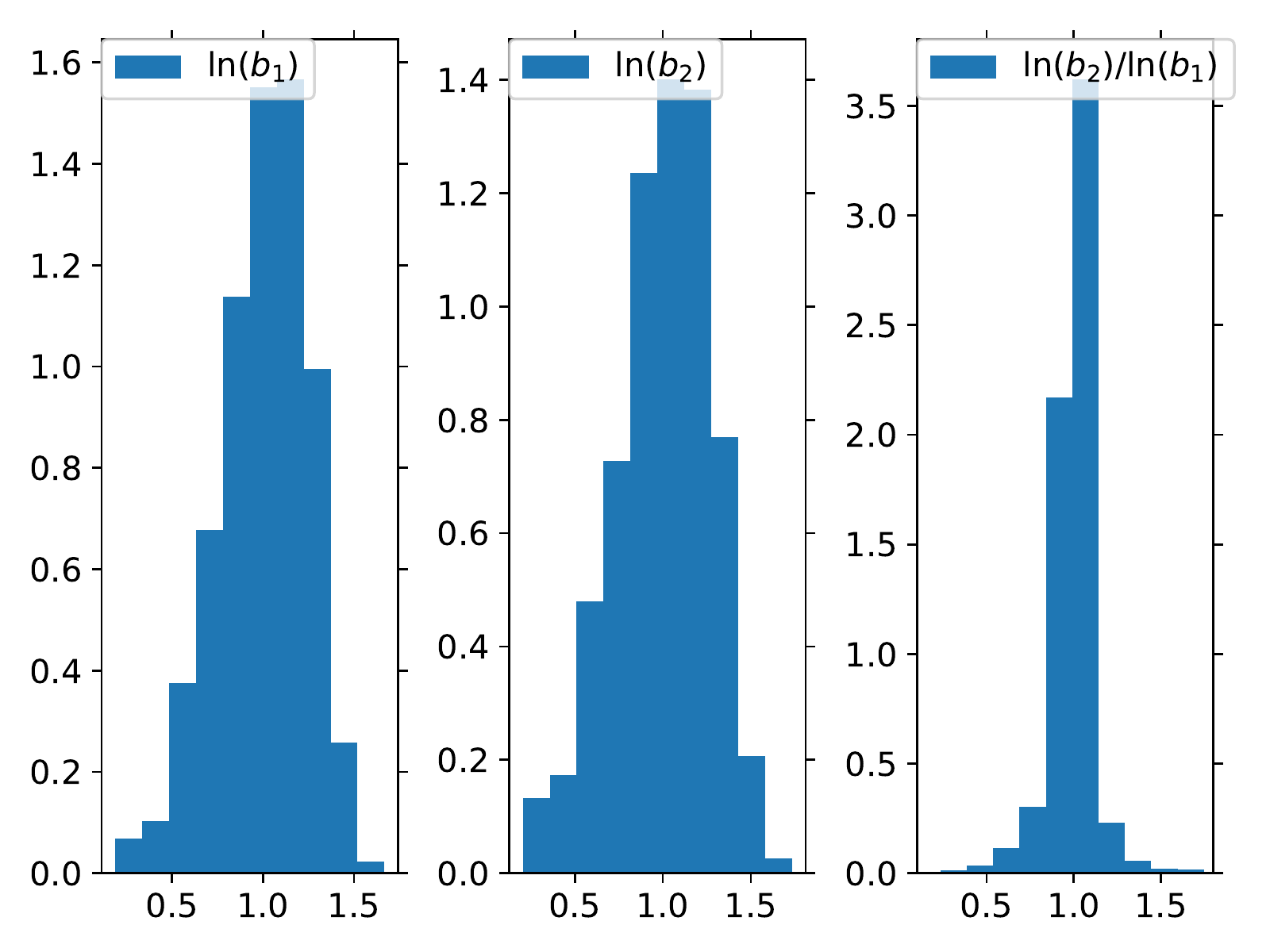}}
  \subfloat{\includegraphics[width = 0.5\textwidth]{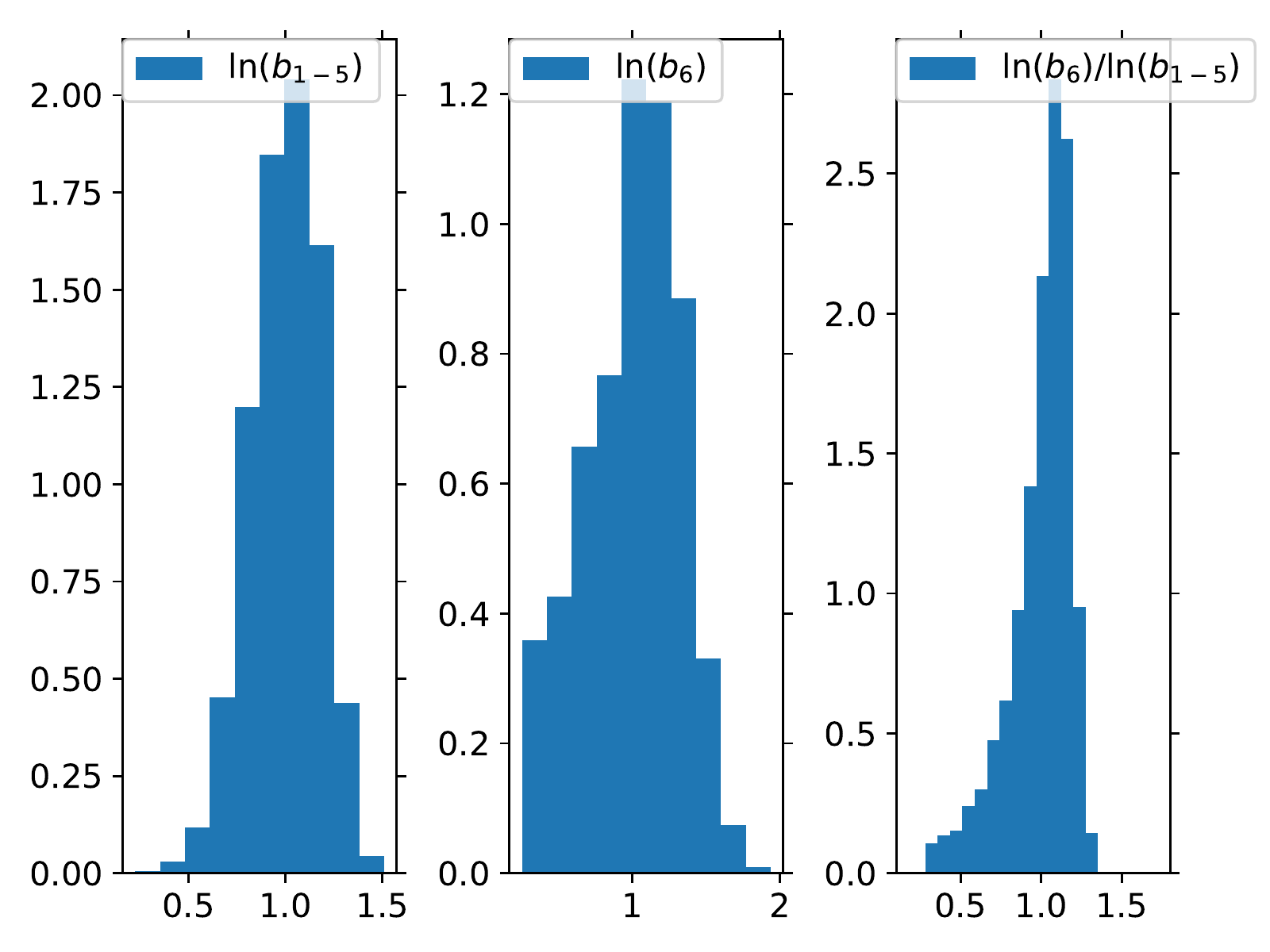}}}
   
  \caption{\textit{Left:} Distribution of the logarithm of measured counts (normalised to the corresponding counts mean values) in energy bin 1 (left panel) and energy bin 2 (middle panel), compared to their ratio (right panel). \textit{Right:} Distribution of the logarithm of counts in energy bin 1 to 5 (left panel) and energy bin 6 (middle panel), compared to their ratio (right panel).
  In all cases, the histograms have been normalised to their mean value to ease the visual comparison.}
\label{fig:correlation}
\end{figure}

In \cite{Calore_2018},  we considered the six energy bins to be fully correlated: In the left panel of Fig.~\ref{fig:correlation}, this is illustrated by comparing the widths of the distribution of counts in the first two energy bins to the width of the distribution of the counts in each bin, separately. Each distribution (individuals and ratios) has been normalised to its mean value for scaling purposes.
The former distributions being significantly narrower than the latter ones confirms that the approximation is reasonable. To assess the impact of this approximation, we study what happens if we relax it for the ``least correlated'' bin. This turns out to be the sixth bin compared to all the others (see right panel in Fig.~\ref{fig:correlation}), whose ratio presents the broadest distribution. In our analysis, we will thus compare the case where the background counts in this bin are profiled independently from all the others, to the case where all bins are treated as fully correlated. These two limiting cases should bracket somewhat the real energy correlation.  
 In one case, a single optimal tuple ($\sigma^*$,$\varsigma^*$) will be adopted; in the other case, two different ones will be chosen for the combination bin 1 to bin 5, and for bin 6, respectively, although, as argued above, this makes little difference to the result. 

\section{Results}\label{results}

\begin{figure}[h]
  \centering{
  \subfloat{\includegraphics[width =0.5\textwidth]{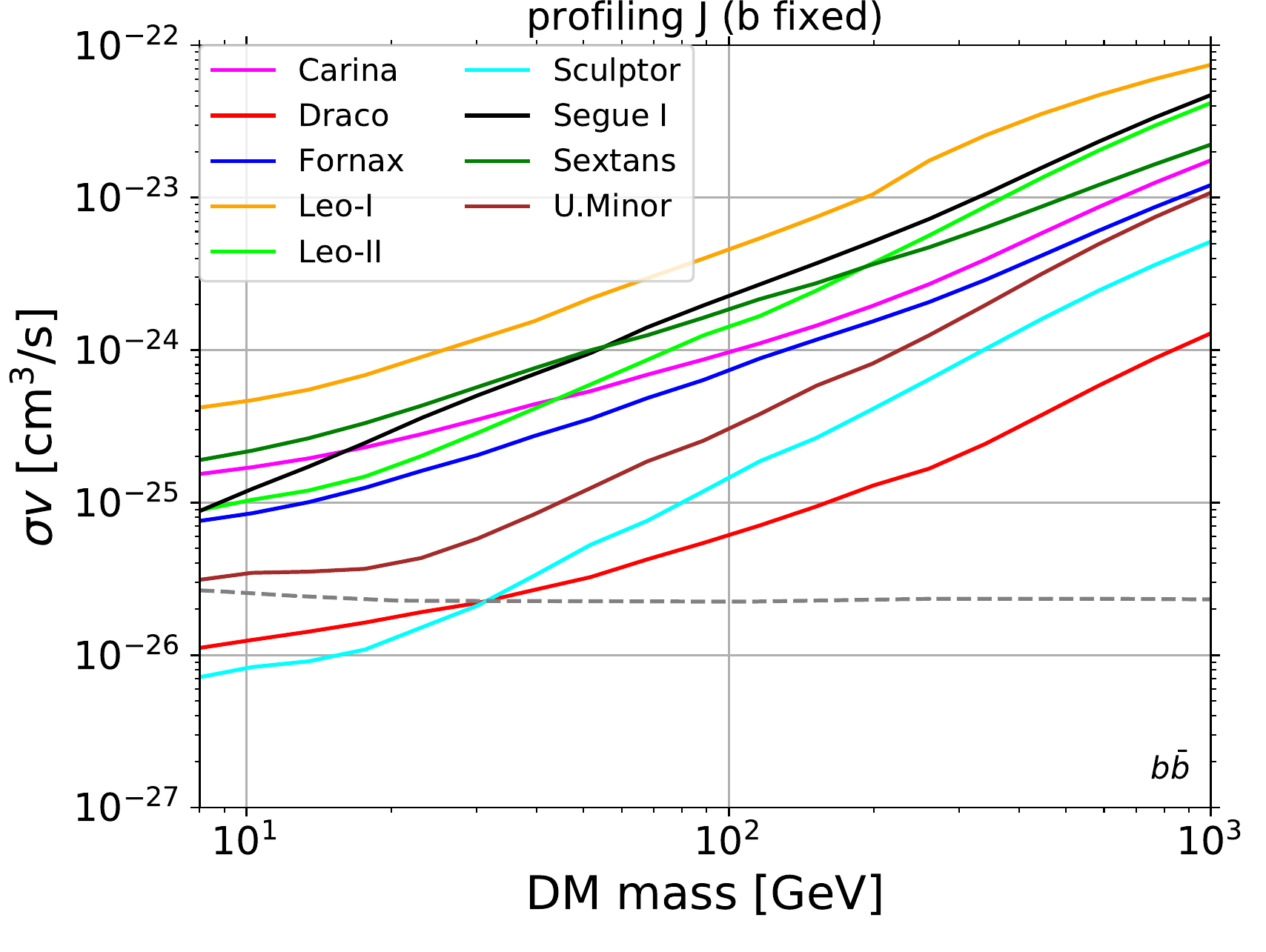}}
  \subfloat{\includegraphics[width =0.5\textwidth]{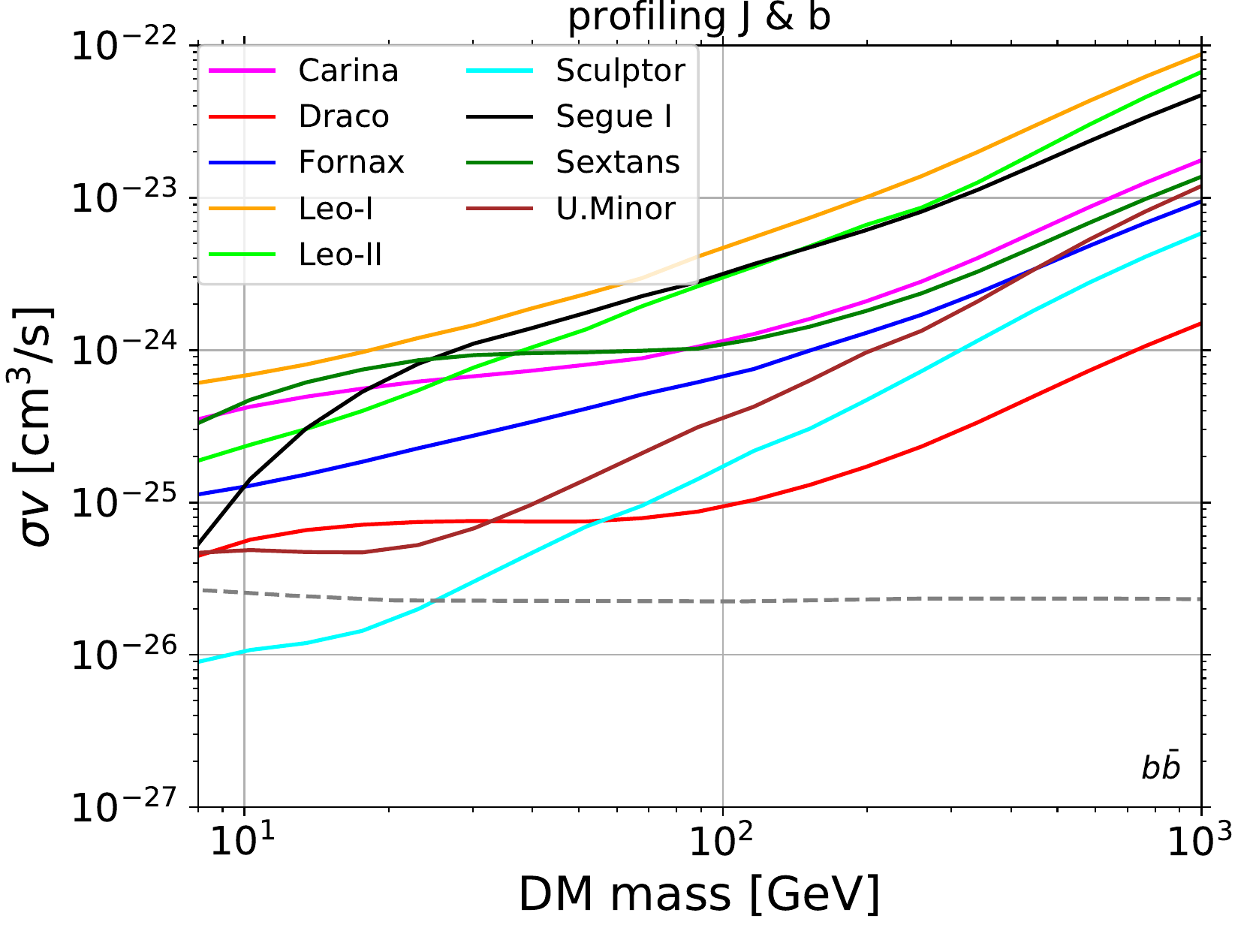}}}
  \caption{\emph{Left}: Bounds from single dSphs when profiling only over \Jf-factor distributions. \emph{Right}: Bounds from single dSphs when profiling  over both \Jf-factor distributions and the background one. The dashed grey line indicates the thermal relic s-wave cross section for generic WIMP models~\cite{Steigman:2012nb}.}
\label{fig:singleJ}
\end{figure}

We present now the results of our analysis. Following the same procedure as in \cite{Calore_2018}, we use the method of profile-likelihood to obtain the limits on the DM annihilation cross section, for a given DM mass. The profiling is done with the \texttt{iminuit} library \cite{iminuit, 1975CoPhC..10..343J} of \texttt{Python3}.  
\newline\newline\noindent
For a single dSph, if profiling over \Jf-factor only (left panel, Fig.~\ref{fig:singleJ}), the results reflect: a) the expectation values of the \Jf-factors; b) the more or less pronounced agreement of the background with the actual counts. Albeit the background estimates are always within 1-2 sigma of the observed counts, some under-prediction (as in the case of Sextans) or over-prediction (as in the case of Sculptor) can easily translate into a factor 10 worse or better bound than naively expected.

Unsurprisingly, given the agreement of the present \Jf-factors analysis with previous ones, in most cases the bounds are similar to results that would be obtained using \Fermi~\Jf-factors, with only  modest degradation or, more often, improvements. Improvements of a few tens of percent in the bounds can follow not only from larger \Jf-factors, but also from  narrower PDF's, which imply less room for degradation when profiling. 

These mild differences are, however, further mitigated when performing a profiling over both $J$ and $b$ (right panel, Fig.~\ref{fig:singleJ}). This is because the profiling over $b$ is at least as important as profiling over $J$, if not even more important, especially for dSphs whose \Jf-factors are very well determined. {\it For most ``classical'' dSphs, we are actually in a situation where the systematic uncertainty in the background dominates over the uncertainty on their \Jf-factors.}

We note the possibility of a non-trivial bound degradation at low DM mass associated to  
our hypothesis of full correlation across energy bins,
when profiling over both $J$ and $b$, taking the case of Draco as an example. 
This can arise when the background matches well the counts at low energy, but {\it overestimates} the counts at high energy. In the particular situation of Draco, the optimisation algorithm prefers to lower the overall level of background in order to fit better the high energy part and, in order to compensate for the newly-created deficit at low energy, to add some light DM annihilation contribution (which does not contribute to the counts at the highest energies).   

The correlation hypothesis is also responsible for the atypical behaviour of Segue I (see Fig.\ref{fig:singleJ}-right). At the location of this dSph, the background overestimates the counts at the lowest energies, whereas it underestimates the counts at the highest energies. Consequently, the optimisation algorithm chooses a non-zero DM contribution to improve the fit at high energies while not exacerbating the overestimate at low energies. This is possible if the DM is heavy enough (since heavy masses do not contribute importantly to the lowest energies).   

In both situations described above, the crucial assumption inducing such effects is the hypothesis of ``rigid'' rescaling of the background in all energy bins. We will shortly come back to this point. We highlight that this effect is simply not noticeable in ``conventional'' analyses of dSphs, where no profiling over the background is performed. 

\begin{figure}[h]
  \centering{
  \subfloat{\includegraphics[width = 0.5\textwidth]{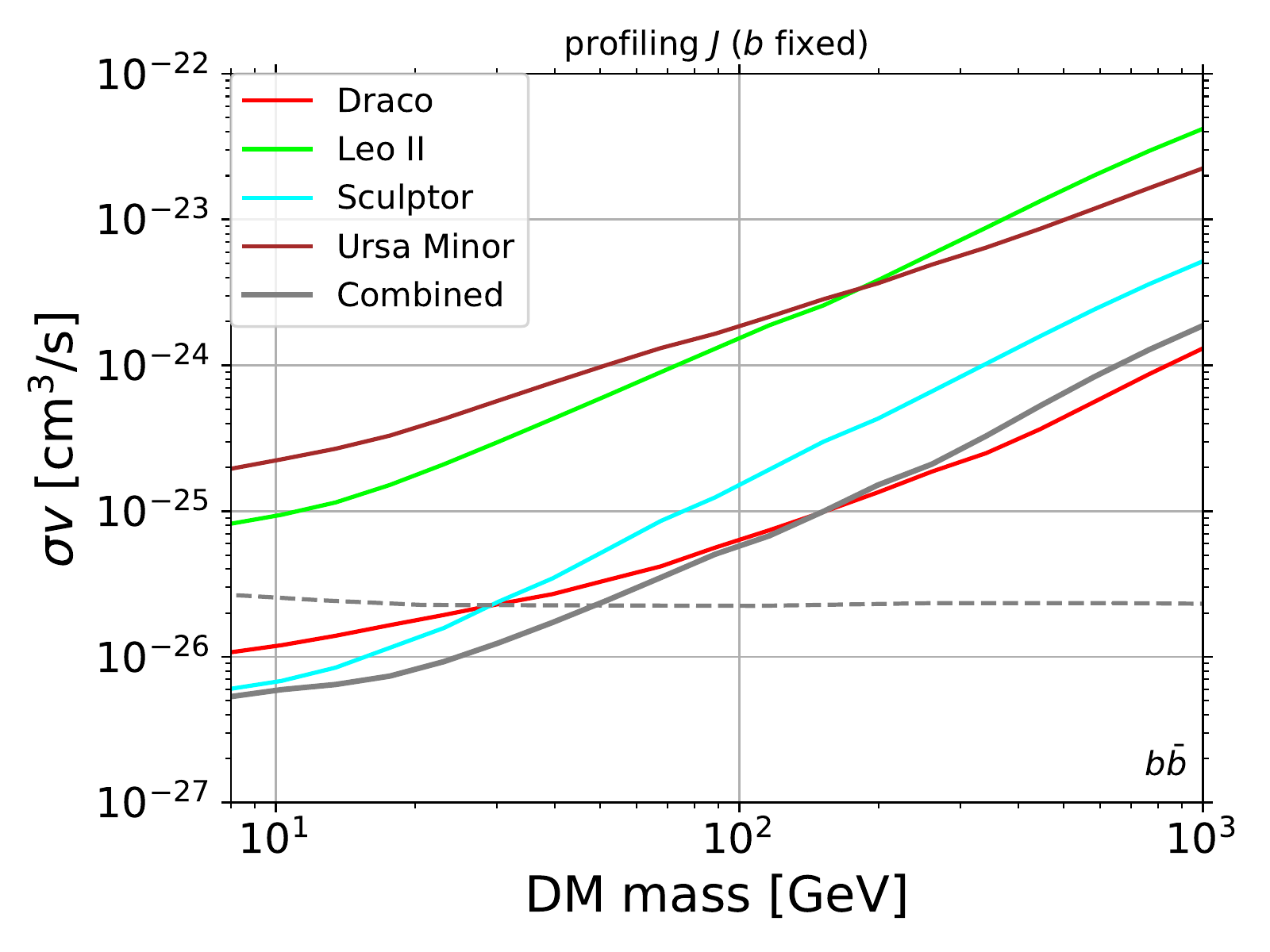}}
  \subfloat{\includegraphics[width = 0.5\textwidth]{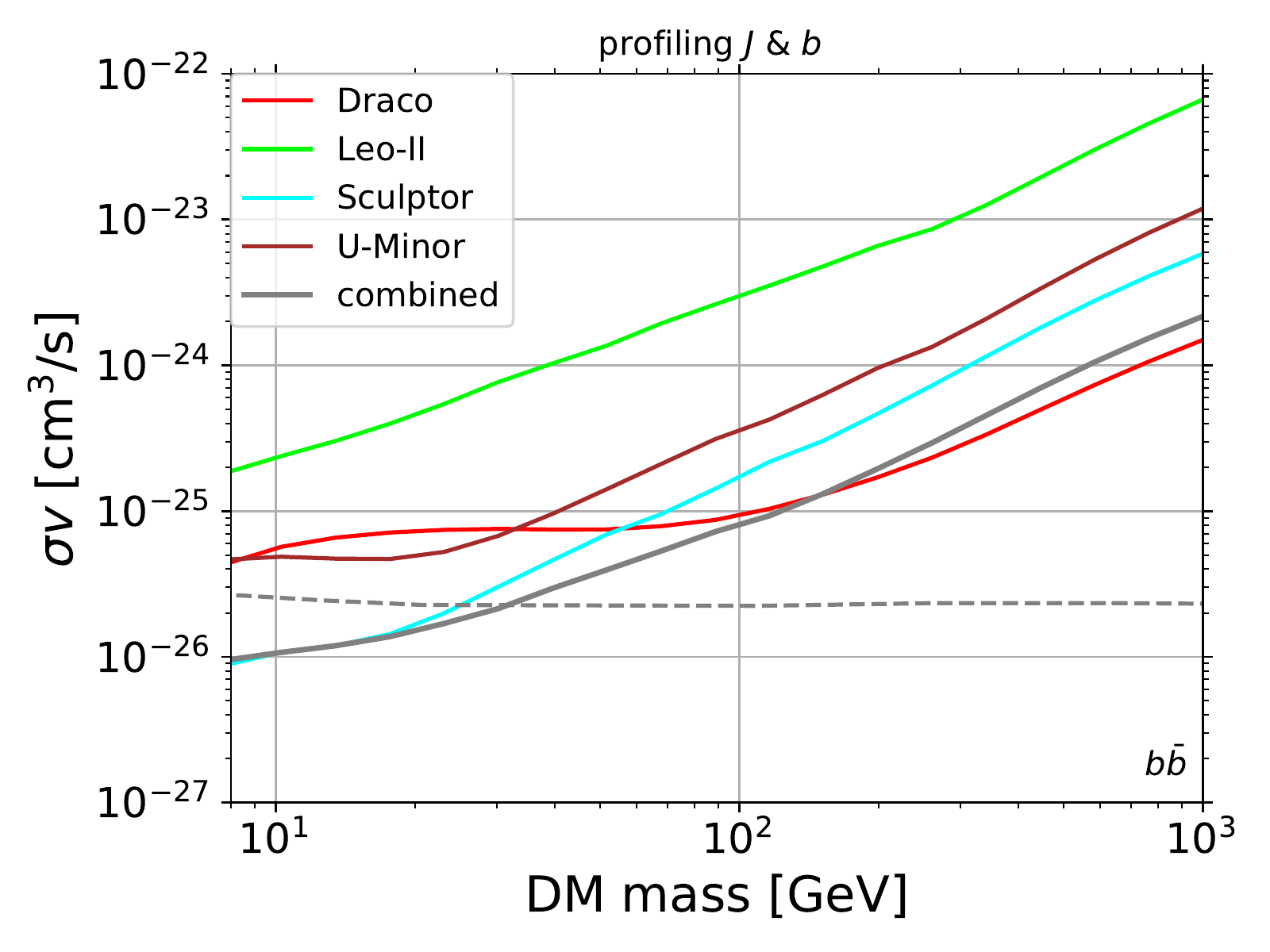}}
  }
  \caption{\emph{Left}: Bound from stacked dSphs when profiling only over \Jf-factor distributions. \emph{Right}: Bound from stacked dSphs when profiling  over both \Jf-factor distributions and the background ones. In both panels, we also show the four best single-dSph of the four dSphs we used in the combined analysis. The dashed grey line indicates the thermal relic s-wave cross section for generic WIMP models~\cite{Steigman:2012nb}.}
\label{fig:case35} 
\end{figure}

When stacking the dSphs (for computational rapidity, we limit ourselves to the four most constraining ones), if profiling over \Jf-factor only (left panel, Fig.~\ref{fig:case35}), we see an improvement of the combined bound over the best four single-dSph limits up to $m\gtrsim 100\,$GeV. The  degradation of the combined bound at high mass follows from the previously mentioned background overestimate at high energy in the case of Draco. When profiling over the background as well (right panel, Fig.~\ref{fig:case35}), the same phenomenon mentioned when commenting the right panel of Fig.~\ref{fig:singleJ} is observable. The stacked bound is slightly worse than the best single-dSph one, but we note that the limit is now much more featureless, resembling the one of the left panel but for a $\sim 30\%$ degradation at low masses.

\begin{figure}[h]
  \centering
  \subfloat{\includegraphics[width = 0.65\textwidth]{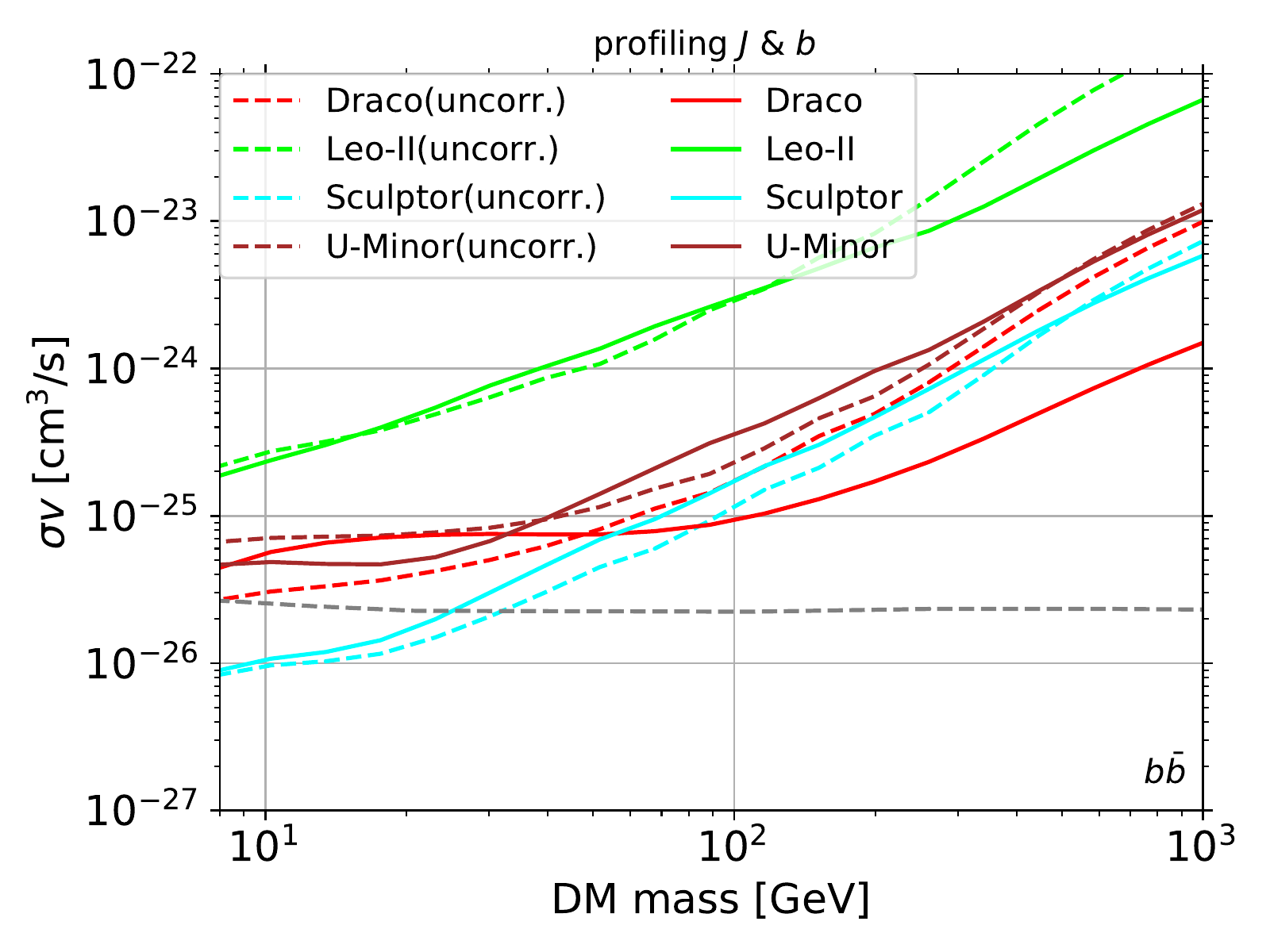}}
  \caption{Comparison between the fully correlated energy-bin hypothesis for the background profiling (solid, benchmark case) and the case where the highest-energy bin (bin 6) is uncorrelated and allowed to be profiled over independently (case ``uncorr.'', dashed curves). The dashed grey line indicates the thermal relic s-wave cross section for generic WIMP models~\cite{Steigman:2012nb}.}
\label{fig:case4_E_dep}
\end{figure}

Given the relevance of the hypothesis of energy bin correlation stressed before, we have studied the impact of relaxing the hypothesis of full energy correlation when profiling over background uncertainties, see Sec.~\ref{bin_correlation}. The results are reported in Fig.~\ref{fig:case4_E_dep}. Whenever there is a good overall agreement of predicted background and observed counts, leaving bin 6 free to be profiled over independently from the first five ones relaxes a bit the constraints, as expected. However, when there is a tension between low-energy and high-energy data, the newly acquired freedom smooths out the previously highlighted non-trivial mass-dependence behaviour of the limits: Given the additional freedom of the highest-energy bin, one can now simply down-scale the counts therein without altering the good agreement at low energies. This exercise is a useful warning against drawing too strong conclusions in presence of an energy-dependent deficit or excess, since the results depend on the global assumptions of the bin-to-bin correlation of the background. This caveat may have been just overlooked in conventional analyses, where a single spectral fit of the background is typically performed in the region of interest and it is presumably dominated by the more numerous counts at low energy.
We stress once again that our ``uncorrelated'' case is used as a toy model to assess the maximum impact of the energy bin-to-bin correlation, with reality lying somewhere in between the two limiting cases. While in principle all the needed information is contained in the data, it is a computational challenge to extract and manipulate multi-dimensional conditional PDF's. This is beyond the scope of the present work, but tackling it would make an interesting and important extension of our data-driven approach in a future analysis.

\section{Discussion and conclusions}
\label{conclusion}

We have presented a fully data-driven analysis of gamma-ray constraints on DM particle annihilation from dSphs with almost ten years of data from the \Fermi-LAT satellite.

For the first time, we have determined the \Jf-factors of classical dSphs without imposing any \emph{a priori} parameterisation of the DM distribution, such as 
the typically adopted Navarro-Frenk-White one, and using latest observational data. The \Jf- (and the analogous $\mathcal{D}$-) factors raw distributions are available upon request for arbitrary integration angles.
In most cases, the \texttt{FreeForm} modelling of DM distribution in dSphs differs very little from a more constraining modelling which imposes a parametric, 
albeit very much general, form for the DM radial profile, i.e. the  \texttt{coreNFWtides} model.
For almost all of the 8 classical dSphs, we found good agreement with the \Jf-factors as determined in previous analyses, but the \Jf-factors' distribution functions are, in our case, much narrower. The reduced width of the distributions, and thus a better determination of the \Jf-factors uncertainties, 
has an impact on the DM limits, where these uncertainties are fully profiled over.
Constraints on DM annihilation are derived by using the empirical \texttt{FreeForm} shape of the \Jf-factors PDFs, but we highlight that a log-normal fit still provides a good approximation to these PDFs. We report the corresponding \Jf-factor mean values and errors in Tab.~\ref{tab:pred}. Ultra-faint dSphs have the potential to strongly constrain the DM parameter space, but, as we showed with the case of Segue I, the data do not yet allow for a precise enough determination of their DM content. Even with our new analysis, the uncertainty in the \Jf-factor determination remains of $\mathcal{O}$(10) times larger than for the ``classical'' dSphs.
In the future, besides the amazing increase in the number of dSphs that we expect in the next decade thanks to the Large Synoptic Survey Telescope (LSST)~\cite{Drlica-Wagner:2019xan}, with the advent of large telescopes like the ELT, 
the prospects to improve the DM distribution reconstruction in those objects are bright~\cite{2015arXiv150104726E}.

As an additional element of novelty, we further developed a purely data-driven method to estimate the background PDF at each dSph position.
With respect to Ref.~\cite{Calore_2018}, we introduced a series of improvements and robustness tests. In particular, we showed that energy-dependent likelihood analyses are crucially sensitive to assumed correlations between energy bins. By removing the correlation of the least correlated bin (namely the highest-energy one), we demonstrated that the hypothesis of perfectly correlated energy bins 
breaks down whenever there is a tension in the data that makes either the low- or high-energy background estimate mis-predict the real counts.
If, for example, the background matches well the counts at low energy but overestimates the counts at high energy, looser bounds at low energy and stronger bounds at high energy arise in the case of the perfectly correlated energy bins, while this effect disappears when the highest-energy bin is treated as fully uncorrelated.
We stress that this behaviour emerged only because our method, contrary to previous analyses in the literature, allows one to properly take into account background uncertainties and profile over them. Since in reality, the different energy bins are neither fully correlated nor fully uncorrelated, our findings highlight the importance, as a next step, of correctly including energy correlations in the background estimates. 

\medskip

%%%%%%%%%%%%%%%%%%%%%%%%%%%%%%%%%%%%%%%%%%%%%%%%%%%%%%%%%%%%%%%%%%%%%%%%%%%%%%%%%%%%%%
\section*{Acknowledgements}
FC and PDS acknowledge support by the Programme National Hautes \'Energies (PNHE) through the AO INSU 2019, grant ``DMSubG" (PI: F. Calore). 
BZ is supported by the Programa Atracci\'on de Talento de la Comunidad de Madrid under grant n. 2017-T2/TIC-5455,
from the Spanish MINECO ``Centro de Excelencia Severo Ochoa'' Programme via grant SEV-2016-0597, and from the Comunidad de Madrid/UAM project SI1/PJI/2019-00294.

%%%%%%%%%%%%%%%%%%%%%%%%%%%%%%%%%%%%%%%%%%%%%%%%%%%%%%%%%%%%%%%
\appendix
\section{Appendix: $\mathcal{D}$-factors}
\label{sec:Dfactors}
We present here the $\mathcal{D}$-factor distributions derived for the eight ``classical'' dSphs studied in this paper and Segue I.
The $\mathcal{D}$-factor is equivalent to the \Jf-factor but for signals of decaying DM and is defined as the integral along the l.o.s. of the DM density in the target of interest:
\begin{equation}
    \mathcal{D}(\Delta \Omega) = \int_{\Delta \Omega} \int_{\rm l.o.s} \rho (l,\Omega') {\rm d}l {\rm d}\Omega' .
    \label{eq:D_factor}
\end{equation}

\begin{figure}[ht!]
  \centering{
  \subfloat{\includegraphics[width=5.2cm]{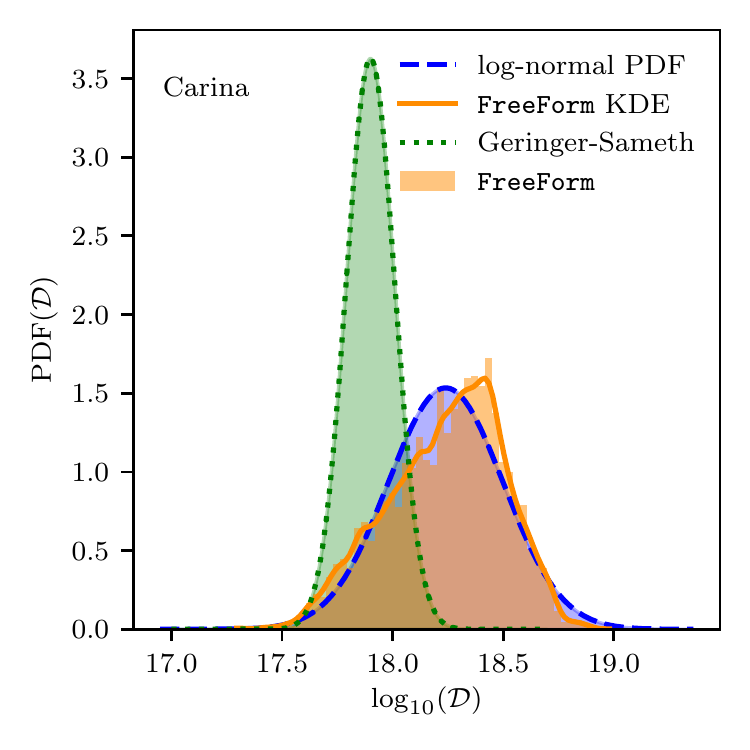}}
  \subfloat{\includegraphics[width=5.2cm]{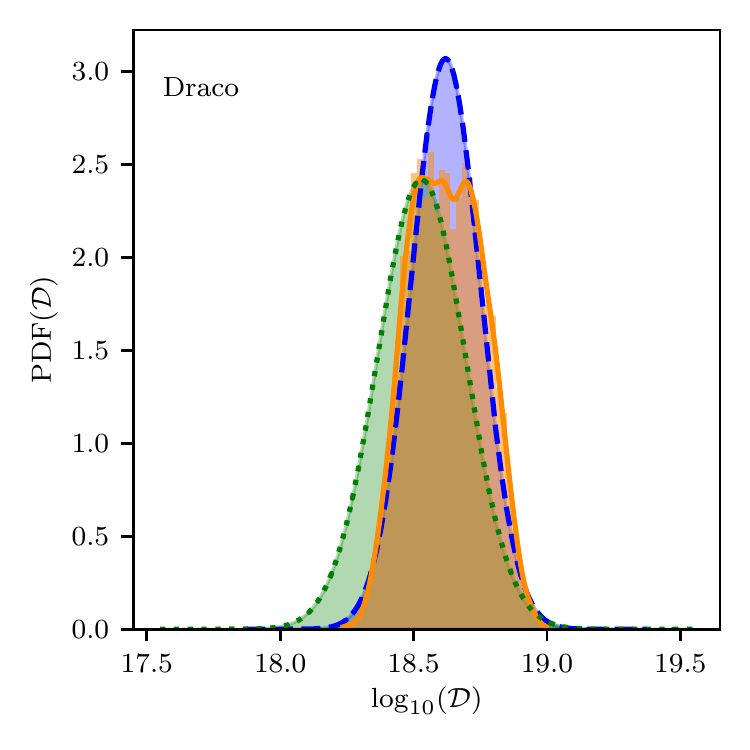}}
  \subfloat{\includegraphics[width=5.2cm]{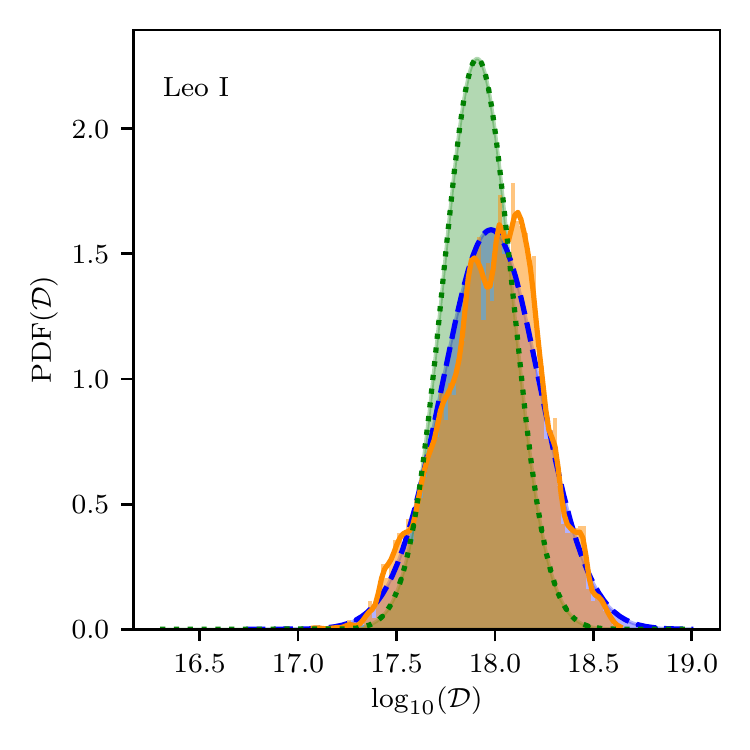}}\\
  \subfloat{\includegraphics[width=5.2cm]{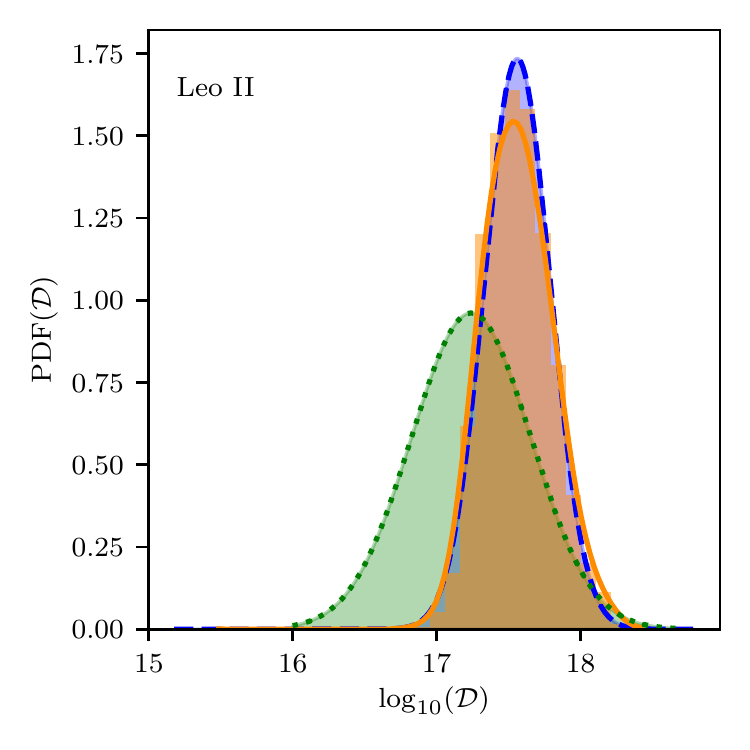}}
  \subfloat{\includegraphics[width=5.2cm]{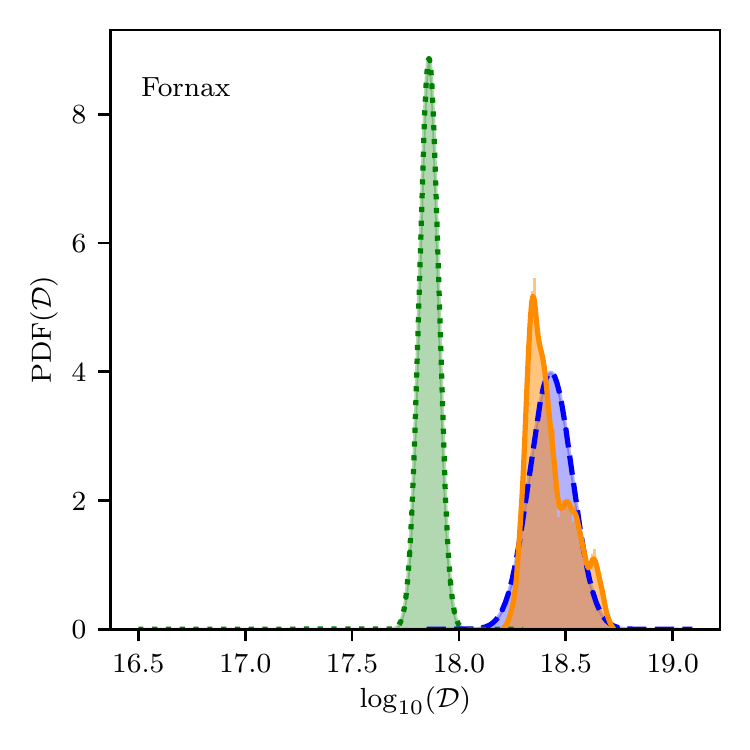}}
  \subfloat{\includegraphics[width=5.2cm]{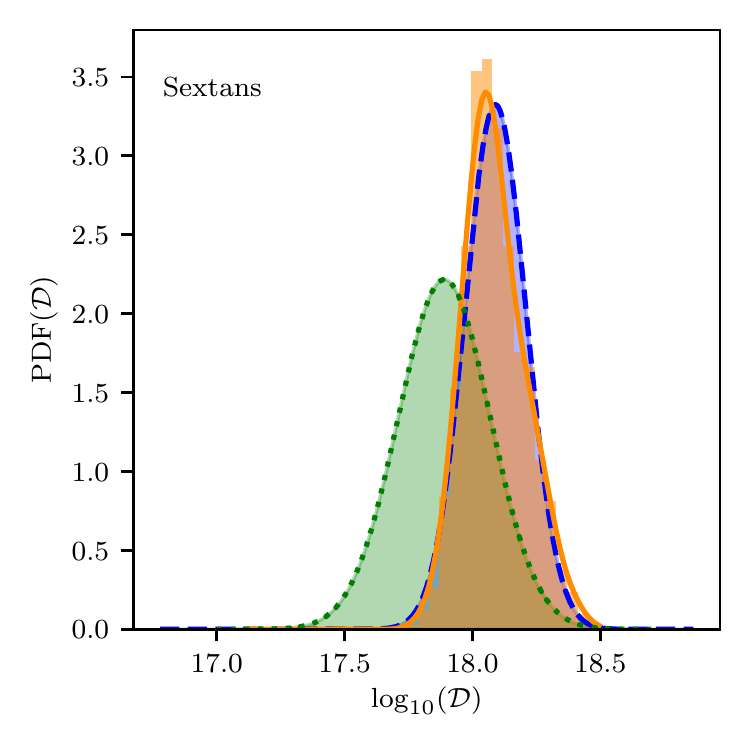}}\\
  \subfloat{\includegraphics[width=5.2cm]{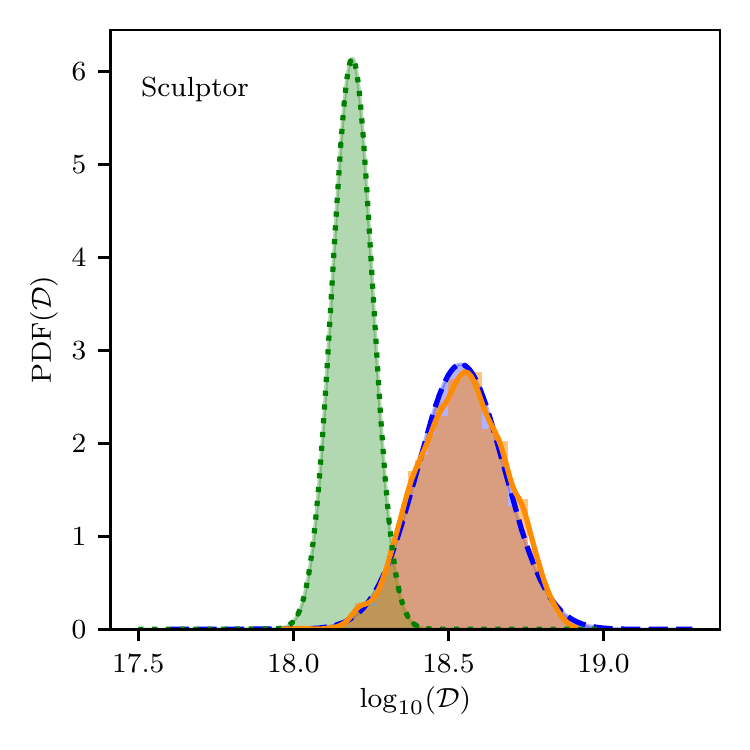}}
  \subfloat{\includegraphics[width=5.2cm]{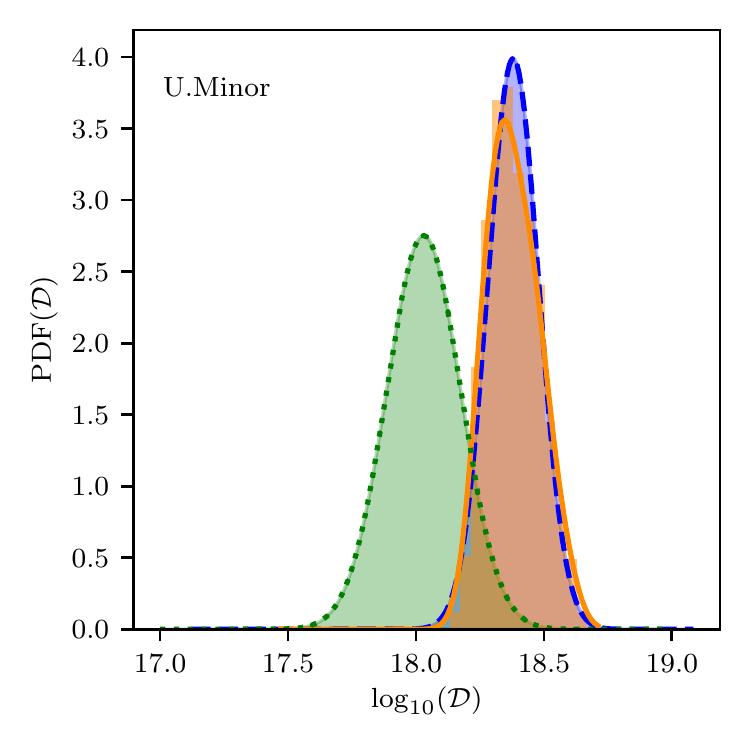}}
  \subfloat{\includegraphics[width=5.2cm]{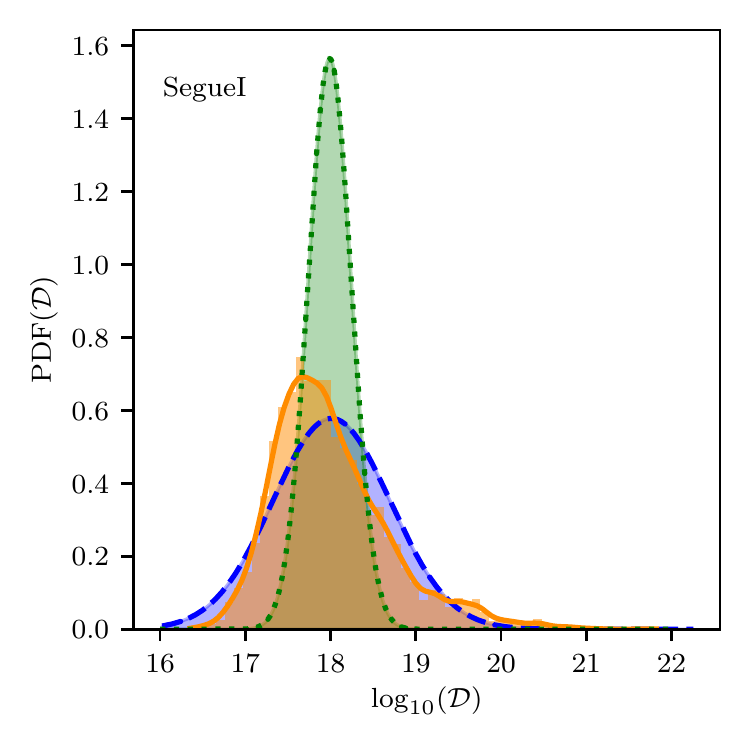}}}
  \caption{New $\mathcal{D}$-factor distributions derived for the 8 ``classical'' dSphs and Segue I for the \texttt{FreeFrom} case (orange histogram, integration angle 0.5$^\circ$) compared with its best-fit log-normal PDF (dashed-blue), and the log-normal distribution from~\cite{Geringer-Sameth:2014yza} (dotted-green, integration angle 0.5$^\circ$) as well.}
\label{fig:D-factors}
\end{figure}

We provide $\mathcal{D}$-factors integrated up to 0.5$^\circ$ or 0.1$^\circ$.
We stress that the integration angle is a very important element to account for in any data analysis and should be chosen consistently with the analysis performed (e.g.~angular resolution, extent of the signal region, etc.). This holds true for both the \Jf- and $\mathcal{D}$-factors.
In Fig.~\ref{fig:D-factors} we show the $\mathcal{D}$-factor distributions for the \texttt{FreeFrom} case, compared with their best-fit log-normal distribution and the results from~\cite{Geringer-Sameth:2014yza}, also provided for an integration angle of 0.5$^\circ$. They are generally in agreement within errors; some differences (for instance due to new data) qualitatively reflect the ones found for \Jf-factors, as already discussed in the main text.

We summarise in Tab.~\ref{tab:Dfact} the results of our log-normal fits to $\mathcal{D}$-factors integrated up to 0.5$^\circ$ and 0.1$^\circ$. 

\begin{table}
\begin{center}
\begin{tabular}{| c | c | c | c |}
\hline
dSph & name & $\log_{10} \mathcal{D}_{0.5^\circ} \pm \sigma^{\mathcal{D}_{0.5^\circ}}$  & $\log_{10} \mathcal{D}_{0.1^\circ} \pm \sigma^{\mathcal{D}_{0.1^\circ}}$ \\
\hline\hline
4 & Carina & $18.24 \pm 0.26$  & $17.15\pm0.11$\\
\hline
6 & Draco & $18.62 \pm 0.13$  & $17.45 \pm 0.06$ \\
\hline
7 & Fornax & $18.43 \pm 0.10$  & $17.22\pm0.05$ \\
\hline
11 & Leo I & $17.98 \pm 0.25$  & $17.04\pm0.09$\\
\hline
12 & Leo II &$17.56 \pm 0.23$ & $16.85\pm0.08$ \\
\hline
17  & Sculptor & $18.54 \pm 0.14$  & $17.35\pm0.07$\\
\hline
18 & Segue I & $18.01 \pm 0.69$& $17.22\pm 0.47$ \\
\hline
19 & Sextans & $18.09 \pm 0.12$& $17.09\pm0.05$ \\
\hline
23 & Ursa Minor &$18.38 \pm 0.69$ & $17.25\pm 0.04$
\\
\hline\hline
\end{tabular}
\caption{Inferred central values and root mean square of the  
$\mathcal{D}$-factor (both in $\log_{10}$ scale) according to our new study. We assume integration angles of 0.5$^\circ$ and  0.1$^\circ$.}
\label{tab:Dfact}
\end{center}
\end{table}

\bibliographystyle{JHEP}
\bibliography{biblio}

\end{document}